\documentclass[twocolumn,showpacs,aps,prl,superscriptaddress]{revtex4}

\usepackage{graphicx}
\usepackage{dcolumn}
\usepackage{amsmath}
\usepackage{epsfig}

\input pubboard/babarsym

\def\pstar     {\ensuremath {p^*}\xspace}

\def\etal {{\it et al.}}

\def\Thetaplus{\Theta(1540)^+}
\def\Ximm{\Xi_5(1860)^{--}}
\def\Xizero{\Xi_5(1860)^0}
\def\Kshort{K^0_s}

\def\Th      {\ensuremath{\Theta^{+}}\xspace}
\def\xmm     {\ensuremath{\Xi_5^{--}}\xspace}
\def\xmn     {\ensuremath{\Xi_5^{0}}\xspace}

\def\xm      {\ensuremath{\Xi^{-}}\xspace}

\newcommand{\BABARPubYear}    {04}
\newcommand{\BABARPubNumber}  {047}

\newcommand{\SLACPubNumber} {10992}

\def\figurebox#1#2#3{%
    \def\arg{#3}%
    \ifx\arg\empty
    {\hfill\vbox{\hsize#2\hrule\hbox to #2{\vrule\hfill\vbox to #1{\hsize#2\vfill}\vrule}\hrule}\hfill}%
    \else
    {\hfill\epsfbox{#3}\hfill}%
    \fi}

\begin{document}


\begin{flushleft}
\babar-PUB-\BABARPubYear/\BABARPubNumber\\
SLAC-PUB-\SLACPubNumber\\
\end{flushleft}

\title{
{\large \bf
Search for Strange-Pentaquark Production in \boldmath{$e^+e^-$} Annihilation 
at \boldmath{$\sqrt{s}=10.58$} \gev 
} }

%
\author{B.~Aubert}
\author{R.~Barate}
\author{D.~Boutigny}
\author{F.~Couderc}
\author{Y.~Karyotakis}
\author{J.~P.~Lees}
\author{V.~Poireau}
\author{V.~Tisserand}
\author{A.~Zghiche}
\affiliation{Laboratoire de Physique des Particules, F-74941 Annecy-le-Vieux, France }
\author{E.~Grauges-Pous}
\affiliation{IFAE, Universitat Autonoma de Barcelona, E-08193 Bellaterra, Barcelona, Spain }
% ...
\collaboration{The \babar\ Collaboration}
\noaffiliation

\date{\today}

\begin{abstract}

We search for strange pentaquark states 
that have been previously reported by other
experiments --- the
$\Theta(1540)^+$, $\Xi_5(1860)^{--}$, and $\Xi_5(1860)^0$ --- in 
123~fb$^{-1}$ of data recorded with the \babar\ detector at
the PEP-II $e^+e^-$ storage ring.
We find no evidence for these states and
set 95\% confidence level upper limits on 
the number of  $\Theta(1540)^+$ and $\Xi_5(1860)^{--}$ pentaquarks produced
per 
$e^+e^-$ annihilation into $q\overline{q}$ and per \Y4S decay.  
For $q\overline{q}$ events these limits are 
about eight and four times lower, respectively, than the 
rates measured for ordinary baryons of similar mass.

\end{abstract}


\maketitle

Ten experimental groups have recently reported narrow 
enhancements near 1540~MeV$/c^2$
in the invariant mass spectra for $n K^+$ or 
$p \Kshort$~\cite{prl91012003,plb572:127,prl91:252001,prl92:032001,diana,svd,hermes,bebc,cosytof,lasttheta}.
The minimal quark content of a state that decays 
strongly to $n K^+$ is $dduu\overline s$;
therefore, these mass peaks have been interpreted as a 
possible pentaquark state, called $\Thetaplus$.
A single experiment (NA49) has reported a narrow 
resonance near 1862~MeV$/c^2$ in the 
invariant mass spectra for $\Xi^- \pi^-$ and $\Xi^- \pi^+$~\cite{prl92:042003}.
The minimal quark content of 
the $\Xi^- \pi^-$ final state is $dssd\overline u$.
Therefore, the latter two mass peaks have also been interpreted as  
possible pentaquark states, named
$\Ximm$ and $\Xizero$, with the latter 
being a mixture of $ussu\overline u$ and $ussd\overline d$.
On the other hand,
a number of experiments that observe large samples of 
strange baryons with mass similar to that of the $\Thetaplus$  
({\it e.g.,} $\Lambda(1520) \to p K^-$)
see no evidence for the $\Thetaplus$~\cite{review};
a number of experiments that observe large samples of the nonexotic 
$\Xi^-$ baryon do not observe the $\Ximm$ or $\Xizero$ states~\cite{review}.

We report the results of inclusive searches for $\Th \!\to \! p \Kshort$,
$\xmm \!\to \!\Xi^-\pi^-$, and $\xmn \! \to \! \Xi^-\pi^+$
in $e^+e^-$ annihilation data, where we expect equal production of the
charge conjugate states; their inclusion is implied throughout this Letter.
The data were recorded with the \babar\ detector~\cite{BABARdetector} 
at the PEP-II asymmetric-energy $e^+e^-$ storage ring 
located at the Stanford Linear Accelerator Center.
The data sample represents an integrated luminosity of 123~fb$^{-1}$
collected at an $e^+e^-$ center-of-mass (CM) energy at or just below 
the mass of the $\Upsilon(4S)$ resonance.

The \babar\ detector is described in detail in Ref.~\onlinecite{BABARdetector}.
We use charged tracks reconstructed in the five-layer silicon vertex tracker
and the 40-layer drift chamber.
The charged-particle momentum resolution is 
$(\sigma(p_T) / p_T)^2 = (0.0013 p_T)^2 + 0.0045^2$,
where $p_T$ is the momentum transverse to the beam axis measured in GeV$/c$.
Particles are identified as pions, kaons, or protons with a combination of 
the energy-loss measured in the two tracking detectors and 
the Cherenkov angles measured in the detector of internally reflected 
Cherenkov radiation.
We use all events accepted by our trigger,
which is  more than 99\% efficient for both $e^+e^- \to q\overline{q}$ 
and $e^+e^- \to \Y4S$ events.

To evaluate the efficiency and mass resolution for reconstructing pentaquarks,
we simulate pentaquark signals with the JETSET~\cite{jetset} Monte Carlo
generator by substituting a particle with the
mass, width, and decay mode of a hypothetical pentaquark for an 
existing baryon already simulated by JETSET. 
We use large control samples of known particles identified 
in data to correct small 
inaccuracies in the performance predicted by the 
GEANT-based~\cite{GEANT} detector simulation.
The invariant-mass resolution for the decay modes studied in this analysis 
ranges from less than 2~MeV$/c^2$ to approximately 8~MeV$/c^2$,
depending on the final state and the momentum of the pentaquark candidate.


We reconstruct \Th candidates in the $p\KS$ 
decay mode, where $\KS \!\!\to\! \pip\pim$.
A sample of \KS candidates is obtained from all pairs of oppositely charged 
tracks we identify loosely as pions (with more than 99\% efficiency 
and [70--90]\% rejection of $K$ and $p$ depending on momentum) 
that pass within 6~mm of each other.
Each candidate is required to have a reconstructed trajectory passing
within 6 mm of the interaction point (IP) in the plane transverse
to the beam direction and within 32~mm of the IP along the 
beam direction,
a positive flight distance, defined as the projection on its
momentum direction of a vector from its point of 
closest approach to the beam axis to its decay point,
and an invariant mass within 10~\mevcc of the nominal \KS mass.
We also require the helicity angle 
$\theta_H$ of the candidate, defined as the angle between the \pip and the
$\pip\pim$ flight directions in the $\pip\pim$ rest frame,
to satisfy $|\cos\theta_H|<0.8$, reducing background from $\Lambda^0$ decays
and photon conversions.

\begin{figure}[hbt]
\begin{center}
\scalebox{0.6}{
\includegraphics[width=14.4cm]{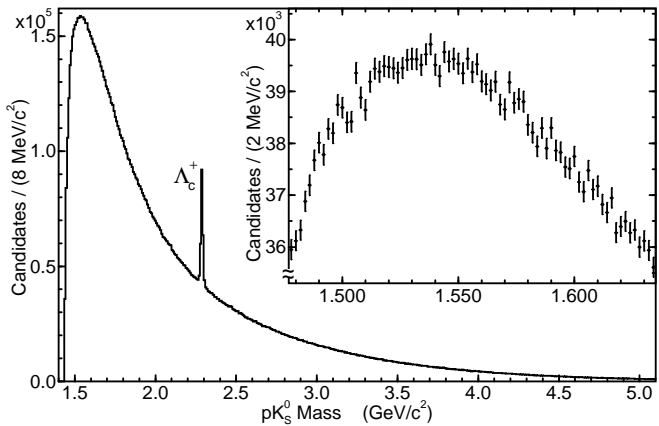}}
\caption{Distribution of the $p\KS$ invariant mass for combinations
  satisfying the criteria described in the text.
  The same data are plotted for the full kinematically allowed $p\KS$ 
  mass range and, in the inset, with statistical uncertainties
  and a suppressed zero on the vertical scale, 
  for the mass range in which the
  $\Thetaplus$ has been reported.
}
 \label{fig:thpmass1}
\end{center}
\end{figure}

We combine these \KS candidates with tracks 
we identify as $p$ or $\overline{p}$ (with [55--99]\% efficiency 
and [95--99]\% rejection of $\pi$ and $K$)
that extrapolate within 15 mm (10 cm) of the IP in the plane 
transverse to (along) the beam direction.
The invariant-mass distribution of $p\KS$ pairs in data is
shown in Fig.~\ref{fig:thpmass1}.  
No enhancement is seen near the mass of the reported $\Thetaplus$
(inset in Fig.~\ref{fig:thpmass1}).
There is a clear peak containing 98,000 entries
at 2285~\mevcc from $\Lambda_c^+ \rightarrow p\KS$,
with a mass resolution below 6~\mevcc.

We consider several additional criteria that might reduce background to a pentaquark signal.
Increasing the required flight distance of the \KS candidates increases
the $\Lambda_c^+$ signal-to-background ratio, but does not reveal any 
additional structure.
We also tried requiring at least one $K^-$ and/or $\bar{p}$ 
candidate in the event.
The $\Lambda_c^+$ signal is
still visible and there is no sign of a pentaquark peak.

To enhance our sensitivity to any production mechanism that
gives a $p\KS$ momentum spectrum in the CM frame
(\pstar) different from that of the background,
we split the data into ten subsamples according
to the value of \pstar for the $p\KS$ candidate.
The ten \pstar ranges are 500~\mevc wide and cover values from 0
to 5~\gevc, the kinematic limit for a particle of mass 1700~\mevcc.
The background is lower at high \pstar, so we are more sensitive to 
mechanisms that produce harder spectra.
There is no evidence of a pentaquark signal in any \pstar range.

We quantify these null results for a \Th mass of 1540~\mevcc.
We fit a signal-plus-background function to the $p\KS$ invariant-mass 
distribution for candidates in each \pstar range.
We use a $p$-wave Breit-Wigner lineshape 
convolved with a resolution function derived from the $\Lambda^+_c$ data and 
simulation. 
The latter is a sum of two Gaussian distributions with a common center and an
overall root-mean-squared-deviation (RMS) 
ranging from 2.5~\mevcc at low \pstar to 1.8~\mevcc at high \pstar;
this is narrower than the $\Lambda^+_c$ resolution due to the proximity of 
1540~\mevcc to $p\KS$ threshold. 
The best upper limit of 8~\mevcc~\cite{diana} on the natural width $\Gamma$ 
of the \Th is larger than 
our $p\KS$ mass resolution, and $\Gamma$ could be very small.
Therefore, we use $\Gamma = 1$~\mevcc and $\Gamma = 8$~\mevcc
in the fit and quote results for each assumed width.
We account for broad structures (known and unknown
resonances, reflections) in the $p\KS$ mass distribution by
using a wide mass range, from threshold to 1800~\mevcc,
and a seventh-order polynomial times a threshold function for the
background shape; seventh is the lowest order giving an
acceptable $\chi^2$.

For the nominal selection criteria, 
we find that in each \pstar range
the fit quality is good and the signal is
consistent with zero.
We consider systematic effects in the fitting procedure by varying 
the signal and background functions and fit range; changes in the
signal yield are negligible compared with the statistical uncertainties.
Varying the mass assumed for the \Th has effects consistent with
expected statistical variations.
The other selection criteria give similar results.
Since the nominal selection results in the
smallest absolute uncertainties after efficiency corrections, we use it
to set upper limits on the production cross section.


We convert the signal yield in each range of \pstar into a cross
section by dividing by the reconstruction and selection
efficiency, the $\KS\to\pip\pim$
branching fraction, the integrated luminosity, and the \pstar range.
If the \Th decays strongly, we expect only two possible decay
modes,  $nK^+$ and $pK^0$, with very similar $Q$ values, 
so we assume ${\cal{B}}(\Th \to p\KS)=1/4$.
The efficiency for the simulated pentaquark signal varies
from 13\% at low \pstar to 22\% at high \pstar.
The efficiency calculation 
is verified by measuring the differential cross section for 
$\Lambda_c^+$ production in the combination of $q\bar{q}$  ($q= d,u,s,c$) 
and \Y4S events 
represented in our data.

\begin{figure}[hbt]
\begin{center}
\scalebox{0.6}{
\includegraphics[width=14.3cm]{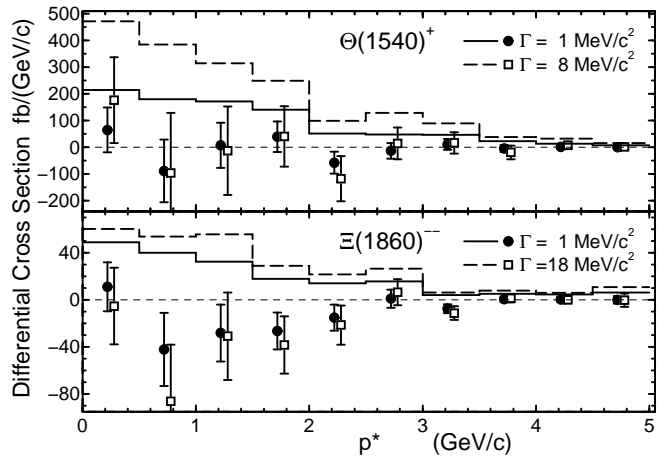}}
\end{center}
\vspace{-0.4truecm}
\caption {
The measured differential production cross sections (symbols) and
corresponding $95\%$ CL upper limits (lines) 
for \Th (top) and \xmm (bottom),
assuming natural widths of $\Gamma=1$~\mevcc (solid) and at the current
experimental upper limit (open/dashed), 
as functions of CM momentum.}
\label{fig:firstul}
\end{figure}

The resulting differential cross sections
are shown for $\Gamma=1$~\mevcc and for $\Gamma=8$~\mevcc
in Fig.~\ref{fig:firstul}.
The error bars include the relative systematic uncertainties on the 
luminosity (1\%) and efficiency (4.9\% dominated by the uncertainties
on track and displaced-vertex reconstruction efficiencies).
We derive an upper limit on the \Th production cross section for each
\pstar range under the assumption that it cannot be negative:
a Gaussian function centered at the measured value with RMS equal to
the total uncertainty is integrated from zero to infinity, and the
point at which the integral reaches 95\% of this total is taken as the limit.
These $95\%$ confidence level (CL) upper limits are
also shown in Fig.~\ref{fig:firstul}.

We derive model-independent upper limits on the total number of pentaquarks
produced per $q\overline q$ event
and per \Y4S\ decay by summing the differential cross section over the 
kinematically allowed \pstar range for $q\bar{q}$ events 
(entire \pstar range)
and for $B$ meson decays (\pstar$<2.5$~\gevc), respectively,
taking into account the correlation in the systematic uncertainty.
The central value and the 95\% CL upper limit on the total 
\Th (plus $\overline{\Theta}^-$) production cross section 
for the \pstar range from 0 to 5~\gevc
are shown in Table~\ref{tab:upperlimits}.
Dividing this limit and the corresponding limit
for the \pstar range from 0 to 2.5~\gevc by the 
cross section for $e^+e^-\rightarrow q\overline q$  
and for $e^+e^-\rightarrow \Y4S$, respectively,
we calculate limits on the number of
pentaquarks per event, given in 
Table~\ref{tab:upperlimits}.
For the maximum width ($\Gamma = 8$~\mevcc),
we obtain a 95\% CL upper limit 
roughly a factor of eight below the 
typical values measured 
for ordinary octet and decuplet baryons of the same mass~\cite{RPP2004}.

\begin{table*}[thb]
\begin{center} 
\caption{
The measured total production cross section and 
95\% CL upper limits (UL) on the cross section and 
yield per event for $\Thetaplus$ and $\Ximm$ pentaquark candidates.
The natural widths $\Gamma=8(18)$~\mevcc refer to the 
upper limits on the widths of the $\Thetaplus$
($\Ximm$), used in the fits.
}
\vspace{0.2cm}
\begin{small} 
\begin{ruledtabular}

\begin{tabular}{ccccccccc}
\vspace{0.1cm} 
          & \multicolumn{2}{c}{total production}
          & \multicolumn{2}{c}{UL on total}
          & \multicolumn{2}{c}{UL on yield per}
          & \multicolumn{2}{c}{UL on yield per}\\
particle  & \multicolumn{2}{c}{cross section (fb)} 
          & \multicolumn{2}{c}{cross section (fb)} 
          & \multicolumn{2}{c}{$q\bar{q}$ event ($10^{-5}$/event)} 
          & \multicolumn{2}{c}{\Y4S decay ($10^{-5}$/event)} \\ [.1cm] 
          & $\Gamma=1$ & $\Gamma=8(18)$
          & $\Gamma=1$ & $\Gamma=8(18)$ 
          & $\Gamma=1$ & $\Gamma=8(18)$ 
          & $\Gamma=1$ & $\Gamma=8(18)$ \mevcc \\ \hline 
$\Th+\overline{\Theta}^-$
& $-$19$\pm$93 & \hspace*{0.3cm} 7$\pm$183 & 171  & 363
                                           & 5.0  &  11 &  18 & 37  \\ 
$\xmm+\overline{\Xi}_5^{++}$
& $-$53$\pm$25 &             $-$93$\pm$ 38 &  25  &  36
                                           & 0.74 & 1.1 & 2.4 & 3.4 \\ 
\end{tabular} 

\end{ruledtabular}
\end{small} 
\label{tab:upperlimits}
\end{center}
\end{table*}


We search, as well, for the reported $\Ximm$ and $\Xizero$
states decaying into a $\Xi^-$ and a charged pion,
where $\Xi^- \to \Lambda^0\pim$ and $\Lambda^0\to p\pim$. 
We reconstruct $\Lambda^0\rightarrow p\pim$ candidates from all pairs of 
charged tracks that satisfy loose proton and pion indentification
requirements and pass within 6 mm of each other.
The $\Lambda^0$ candidate must have a positive flight distance from the IP
and an invariant mass within 10~\mevcc 
of the nominal $\Lambda^0$ mass.
These $\Lambda^0$ candidates are combined with an additional
negatively charged track 
passing loose pion identification requirements to form $\Xi^-$ candidates,
which are required to form a good vertex, to have a
positive flight distance from the IP, 
and to have an invariant mass within 20~\mevcc of the nominal $\Xi^-$ mass.
The flight distance of the $\Lambda^0$ candidate from the 
$\Lambda^0\pim$ vertex is required to be positive.
This selection yields 290,000 $\Xi^-$ candidates with a peak 
signal-to-background ratio of 23 in the $\Lambda^0 \pim$ mass distribution.
Finally, we combine the $\Xi^-$ candidates with an additional charged track 
consistent with coming from the IP and passing 
loose pion identification requirements.
The cosine of the angle between the reconstructed $\xm$ trajectory, 
extrapolated back to the IP, and the additional track is required to be less 
than 0.998.
This last requirement is especially important, 
since the $\xm$ is charged and has a long lifetime;
if it has a long flight distance, it can produce a reconstructed track
that, if combined with itself, forms a false peak in the 
invariant-mass distribution.
The reconstruction efficiency for the simulated pentaquark signal varies 
from 6.5\% at low \pstar to 12\% at high \pstar.

\begin{figure}[hbt]
\begin{center}
\scalebox{0.6}{
\includegraphics[width=14.4cm]{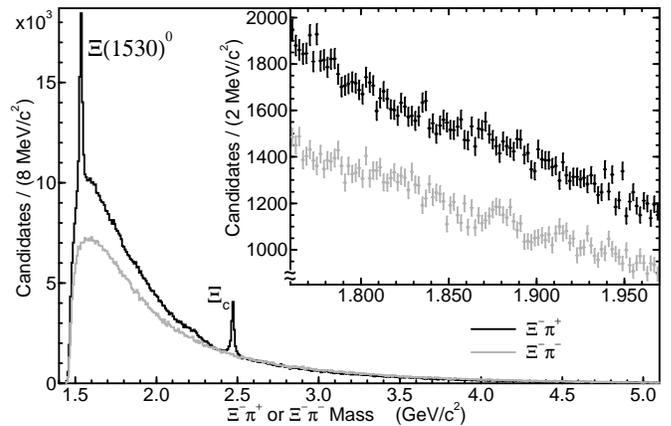}}
\end{center}
\vspace{-0.4truecm}
\caption {$\xm \pip$ (black) and $\xm \pim$ (gray) invariant-mass 
  distributions.
  The same data are plotted for the full kinematically allowed $\xm\pi^\pm$ 
  mass range and, in the inset, with statistical uncertainties
  and a suppressed zero on the vertical scale, 
  for the mass range in which the
  $\Ximm$ and $\Xizero$ have been reported.
} 
\label{fig:mxipipm}
\end{figure}

The invariant-mass distributions for $\xm \pim$ and for $\xm \pip$ 
combinations are shown in Fig.~\ref{fig:mxipipm}.
In the $\xm \pip$ mass spectrum, we see clear peaks for the
$\Xi(1530)^0$ and $\Xi_c(2470)^0$ baryons, but no other structure
is visible.
There are no visible narrow structures in the $\xm \pim$ mass spectrum.

As in the \Th search, we divide the $\xm \pim$ candidates into ten subsamples
according to the \pstar value of the candidate.
We find no sign of a pentaquark signal for any range of \pstar.
We fit a signal-plus-background function to the $\xm\pim$
invariant mass distribution, for each \pstar range.
Here no broad resonances or reflections are evident, and 
we perform simpler fits over a $\xm\pim$ mass range 
from 1760 to 1960~\mevcc using 
a linear background function.
The resolution function is derived from the $\Xi(1530)^0$ and 
$\Xi_c(2470)^0$ signals in 
data and simulation, and is described by a Gaussian 
function with an RMS of 8~\mevcc. 
For the Breit-Wigner width we consider two possibilities, 1 and 18~\mevcc, 
corresponding to a very narrow state and the 
experimental upper limit on the $\xmm$
width~\cite{prl92:042003}, respectively.
We fix the \xmm mass to 1862~\mevcc.
In all ranges of \pstar, the
signal is consistent with zero.
Systematic uncertainties on the fitting procedure are again found to be
negligible compared with the statistical uncertainties, and variations
of the \xmm mass and selection criteria give consistent results.


We convert the measured yields for the $\xmm \to \xm\pim$ decays 
into cross sections as described above for the \Th.
The efficiency determined from simulation is verified by measuring the 
differential cross section for the observed $\Xi(1530)^0$ signal.
The average relative systematic uncertainty on the efficiency is 6.2\%
with a slight \pstar dependence, and
is larger than that for the $p\KS$ mode because there are two displaced
vertices and more particles in the final state.
We have used a $\xm\pim$ 
branching fraction of one-half for purposes of calculating cross
sections and limits, under the assumption that the two-body modes
$\xm\pim$ and $\Sigma^-K^-$ dominate and have similar branching fractions.

The measured cross section and 95\% CL upper limits for 
$\xmm$ (plus $\overline{\Xi}_5^{++}$) production 
are shown in Fig.~\ref{fig:firstul} and Table~\ref{tab:upperlimits}.
For $\Gamma = 18$~\mevcc, the limit  on the total production rate per
$q\bar{q}$ event is roughly a factor of four below the typical values measured 
for ordinary octet and decuplet baryons of the same mass~\cite{RPP2004}.

We perform a similar search for $\xmn \!\!\to \!\!\xm\pip$, 
finding no signal in any \pstar bin.
Since many decay modes are kinematically accessible to such a state
with a mass of $~\sim$1862~\mevcc
and the branching fraction is unknown a priori,
we omit this state from Table~\ref{tab:upperlimits}
and express our upper limit on the total production of \xmn and
$\overline{\Xi}_5^0$
per $q\overline{q}$ event as
$0.8\times10^{-5}/{\cal{B}}(\xmn\to \xm\pip)$,
at the 95\% CL.


In summary, we have performed a search for the reported pentaquark states
$\Thetaplus$, $\Ximm$, and $\Xizero$ in $e^+e^-$ annihilations.
We observe large signals for known baryon states but 
no excess at the measured mass values for the pentaquark states. 

We are grateful for the excellent luminosity and machine conditions
provided by our \pep2\ colleagues, 
and for the substantial dedicated effort from
the computing organizations that support \babar.
The collaborating institutions wish to thank 
SLAC for its support and kind hospitality. 
This work is supported by
DOE
and NSF (USA),
NSERC (Canada),
IHEP (China),
CEA and
CNRS-IN2P3
(France),
BMBF and DFG
(Germany),
INFN (Italy),
FOM (The Netherlands),
NFR (Norway),
MIST (Russia), and
PPARC (United Kingdom). 
Individuals have received support from CONACyT (Mexico), A.~P.~Sloan Foundation, 
Research Corporation,
and Alexander von Humboldt Foundation.

\end{document}